\documentclass[11pt]{article}
\baselineskip
\pdfoutput=1
\usepackage{amsmath}
\usepackage{amssymb}
\usepackage{graphicx}
\usepackage{color}
\usepackage{cite}
\usepackage{collref}
\usepackage{tabls}
\usepackage{hyperref}
\DeclareMathOperator{\Orth}{O}
\DeclareMathOperator{\erf}{erf}

\newcommand{\s}{{\bf{s}}}
\newcommand{\Tc}{T_{\mbox{\tiny{c}}}}

\newcommand{\redchisq}{\chi^2_{\tiny\mbox{red}}}
\newcommand{\eq}[1]{\begin{equation}\label{#1}}
\newcommand{\en}{\end{equation}}
\newcommand{\eqar}[1]{\begin{eqnarray}\label{#1}}
\newcommand{\enar}{\end{eqnarray}}

\begin {document}

\begin{titlepage}

\begin{center}
{\Large\bf Topological excitations in statistical field theory at the upper critical dimension}
\end{center}
\vskip1.3cm
\centerline{Marco~Panero\footnote{\href{mailto:marco.panero@unito.it}{{\tt marco.panero@unito.it}}} and Antonio~Smecca\footnote{\href{mailto:antonio.smecca@unito.it}{{\tt antonio.smecca@unito.it}}}}
\vskip1.5cm
\centerline{\sl Department of Physics, University of Turin and INFN, Turin}
\centerline{\sl Via Pietro Giuria 1, I-10125 Turin, Italy}
\vskip1.0cm

\setcounter{footnote}{0}

\begin{abstract}
\noindent 
We present a high-precision Monte~Carlo study of the classical Heisenberg model in four dimensions. We investigate the properties of monopole-like topological excitations that are enforced in the broken-symmetry phase by imposing suitable boundary conditions. We show that the corresponding magnetization and energy-density profiles are accurately predicted by previous analytical calculations derived in quantum field theory, while the scaling of the low-energy parameters of this description questions an interpretation in terms of particle excitations. We discuss the relevance of these findings and their possible experimental applications in condensed-matter physics.
\end{abstract}

\end{titlepage}

\section{Introduction}
\label{sec:introduction}

Lattice models of interacting vector spins with global $\Orth(N)$ symmetry have many important theoretical as well as experimental applications. Considering only nearest-neighbor interactions, their Hamiltonian can be written as~\cite{Stanley:1968do}
\eq{Hamiltonian}
\mathcal{H} = - J \sum_{\langle x, y \rangle} \s(x) \cdot \s(y),
\en
where $\s(x)$ is an $N$-component real vector of unit length, defined on the site $x$ of a regular Euclidean lattice in $D$ dimensions, the summation runs over all distinct pairs of nearest-neighbor sites, and $J>0$ corresponds to ferromagnetic coupling. As particular cases, eq.~(\ref{Hamiltonian}) includes the self-avoiding random-walk model (for $N=0$), the Ising model (for $N=1$), the XY model (for $N=2$), the Heisenberg model (for $N=3$), a toy model for the Higgs sector in the Standard Model of elementary particle physics (for $N=4$), and the spherical model (for $N \to \infty$). In the present work, we study the Heisenberg model in $D=4$, where a long-range-order phase is known to exist at sufficiently low temperatures~\cite{Froehlich:1981ib} and, $D=4$ being the upper critical dimension, the critical exponents at the phase transition that separates the low-temperature phase from the disordered, high-temperature one, are equal to their mean-field values, up to logarithmic corrections~\cite{Kenna:2004cm, Bauerschmidt:2014sl}. Nevertheless, the model encodes non-trivial dynamics: in particular, in this work we study a class of non-local, finite-energy excitations (topological defects) in the low-temperature, broken-symmetry phase and show, by numerical Monte~Carlo simulations, that their properties can be successfully predicted using quantum-field-theoretical tools in a continuum formulation of the model~\cite{Delfino:2014rja}. For lower-dimensional systems, the theoretical expectations derived in that work are consistent with exact results for the field theory describing the continuum limit of the Ising model in two dimensions~\cite{Berg:1978sw, Delfino:2003yr} and with numerical results for the XY model in three dimensions~\cite{Delfino:2018bff}. Here, for the first time, we provide evidence supporting these predictions, at the quantitative level, also in a four-dimensional Euclidean spacetime, where these excitations have some analogies with monopole-like states (even though this terminology should be taken with a grain of salt, since the model that we are considering has no gauge symmetry nor gauge fields). As will be shown below, however, we also find that the scaling properties of the parameters of the quantum-field-theory model are different from what was originally conjectured in ref.~\cite{Delfino:2014rja}, and challenge a possible interpretation of the topological configurations discussed in this work as physical particles.

\section{Computation setup}
\label{sec:computation_setup}

We consider the system defined by the Hamiltonian introduced in eq.~(\ref{Hamiltonian}), working on a four-dimensional isotropic, hypercubic lattice of spacing $a$. We denote the linear extent of the system in each of the three spatial directions as $L$ (with spatial coordinates ranging from $-L/2$ to $L/2$), while $R$ is the size of the system in the Euclidean-time direction. We also denote the system temperature as $T$ and introduce the reduced temperature $t=(T-\Tc)/\Tc$, where $\Tc$ is the critical temperature. In this work, we use the most recent estimate of the critical temperature, $\Tc/J=2.19879(2)$, which was obtained in ref.~\cite{Lv:2019cke} through a sophisticated finite-size-scaling analysis of the Binder cumulant associated with the magnetization.

In our Monte~Carlo simulations, Markov chains of vector-field configurations are generated by a combination of local heat-bath~\cite{Creutz:1980zw, Miyatake:1986ot, Berg:2004fd} and overrelaxation~\cite{Adler:1981sn} updates. For a subset of our runs, we also use non-local, single-cluster updates~\cite{Wolff:1988uh}. Our production runs were executed in perfectly parallel workloads on machines equipped with Intel Xeon Skylake processors. 

We study the system both in the high-temperature and in the low-temperature phases. In the high-temperature phase, periodic boundary conditions are assumed in the four directions. We consider the zero-spatial-momentum spin operators
\eq{zero-momentum_operator}
{\bf{S}}(x_0) = \frac{a^3}{L^3} \sum_{x_1,x_2,x_3} \s(x)
\end{equation}
and extract the mass $m$ of the lightest physical state by fitting the two-point correlation function
\eq{correlator}
G(\tau,R) = \frac{a}{R}\sum_{x_0} {\bf{S}}(x_0) \cdot {\bf{S}}(x_0+\tau)
\en
to the functional form
\eq{expected_correlator_form}
G(\tau,R) = A \left\{ \exp \left( -m \tau \right) + \exp\left[-m \left( R - \tau \right) \right] \right\}
\en
for sufficiently large values of $\tau$. In the symmetric, high-temperature phase, we run numerical simulations on lattices of sizes $(L/a)^4$ ranging from $40^4$ to $104^4$ and for reduced temperatures in the range $0 \lesssim t \lesssim 0.06$.

Conversely, in the low-temperature phase we impose boundary conditions enforcing the existence of a ``monopole-like'' spin configuration. This is done by taking advantage of the topological nature that this type of excitations have in the continuum: they can be constructed by mapping the spatial boundary of the continuum system at fixed Euclidean time, which has the topology of the $S^2$ sphere, to the manifold of (classical) vacua of the theory in the broken-symmetry phase, which is also the $S^2$ sphere. The simplest non-trivial mapping of this type, enforcing the existence of a single, isolated, monopole, is the one that, for the points at the spatial boundary of the system, identifies the direction of $\s(x)$ with the direction of the spatial component of $x$ (with respect to the center of the system). We also impose the boundary conditions identifying the direction of $\s(x)$ with the direction of the spatial component of $x$ for \emph{all} points at the ``initial'' ($x_0=-R/2$) and ``final'' ($x_0=R/2$) Euclidean times. Then, up to an overall normalization, the partition function of the system $Z$ can be identified with the probability amplitude for the monopole propagation from $x_0=-R/2$ to $x_0=R/2$. We determine the $i$-th component of the magnetization $\langle {\s}_i (x_i) \rangle$ and the energy density profile $\langle \varepsilon (x_i) \rangle$ along the $i$-th spatial axis through the center of the system (so that $\varepsilon (x_i)$ is proportional to the scalar product of $\s$ with the sum of the spins on the nearest-neighbor sites, and is normalized to $1$ for a uniform field configuration). To increase statistics, we average over the three spatial axes.

\section{Results}
\label{results}

\begin{figure*}[]
\centerline{\includegraphics[height=0.28\textheight]{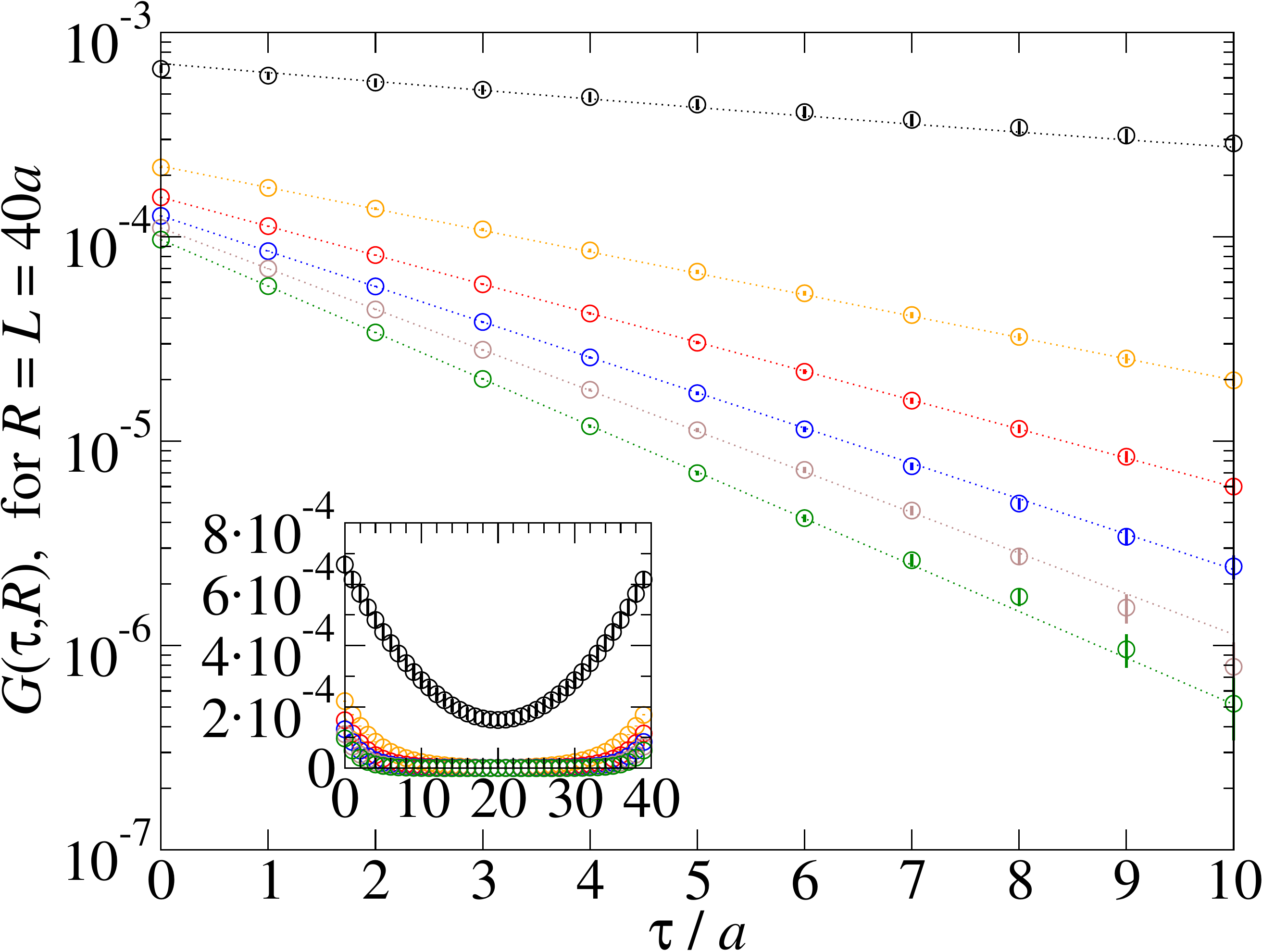} \hfill \includegraphics[height=0.28\textheight]{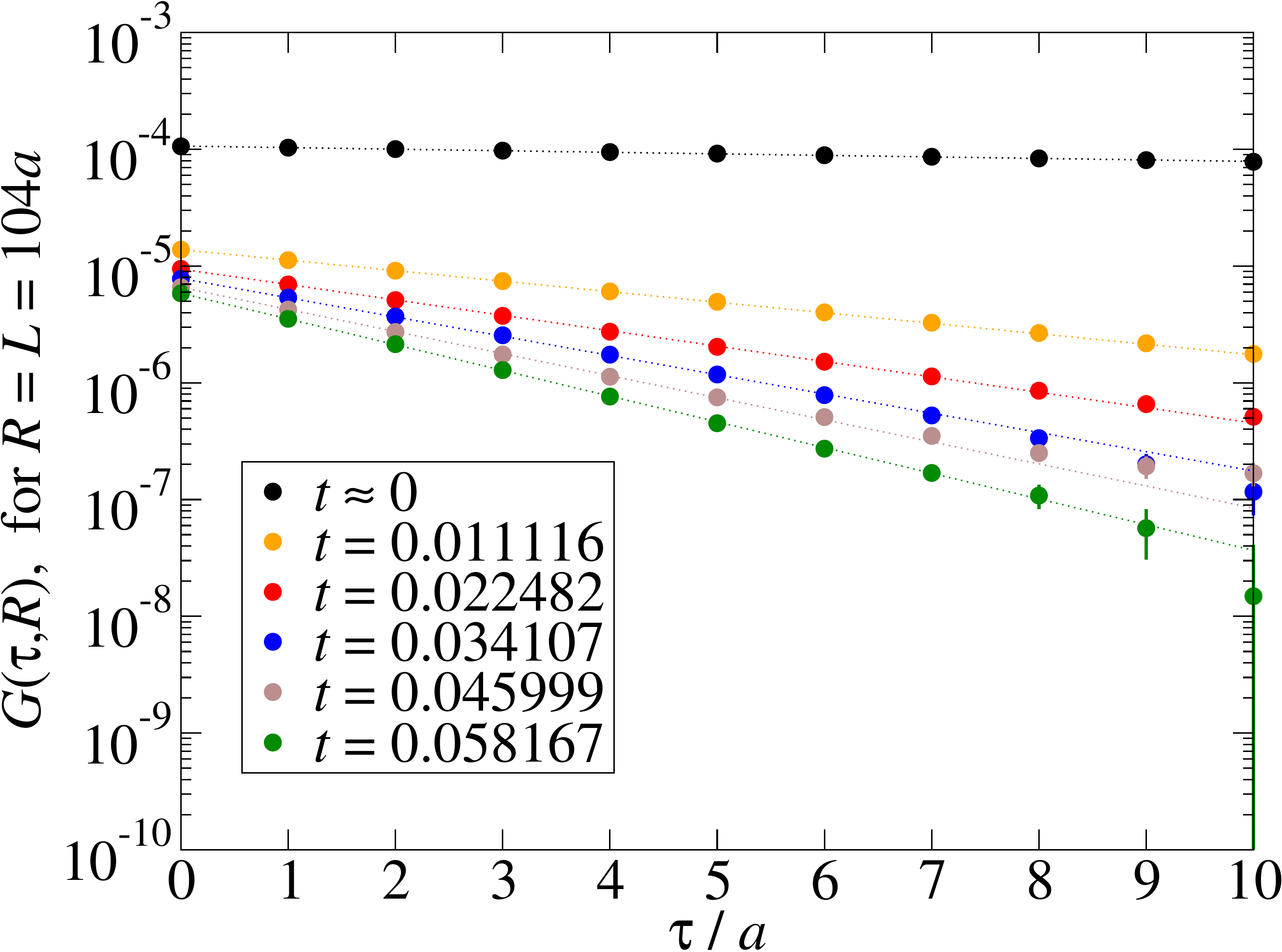}}
\caption{Euclidean-time-separation dependence of the two-point correlation function of zero-momentum operators defined in eq.~(\ref{zero-momentum_operator}), on lattices of size $L^4=(40a)^4$ (left-hand-side panel, open symbols) and $L^4=(104a)^4$ (right-hand-side panel, full symbols), for different values of the reduced temperature $t=(T-\Tc)/\Tc$, denoted by different colors, in the symmetric phase. The curves are obtained from two-parameter fits of our Monte~Carlo simulation results to the expected functional form in eq.~(\ref{expected_correlator_form}). The inset in the left-hand-side panel shows the results for the correlation function for all values $0 \le \tau <L$.} \label{fig:spin_correlators}
\end{figure*}

The results of our calculations in the symmetric phase confirm the estimate of the critical temperature reported in ref.~\cite{Lv:2019cke}. As an example, figure~\ref{fig:spin_correlators} shows our results for the two-point, zero-momentum correlation function defined in eq.~(\ref{correlator}) on the smallest ($L^4=(40a)^4$, left-hand-side panel) and on the largest ($L^4=(104a)^4$, right-hand-side panel) lattices. In particular, the results of our Monte~Carlo calculations close to $\Tc$ (shown as black symbols) indicate that the mass of the lightest physical state that propagates in the theory, vanishes in the thermodynamic limit.
This statement is made more quantitative by the fit results for $ma$ for $T \approx \Tc$, which are reported in table~\ref{tab:m_results}. The estimated uncertainties on all results are always of the order of $1\%$ or less. We note that, while essentially all of our data for $L/a \gtrsim 50$ appear to be consistent with a simple, $1/L$-decay and with the expected exactly zero asymptotic value for $L/a\to \infty$, it is worth mentioning that the precise form of finite-size corrections in this model is not yet a completely settled issue (see, e.g., the discussion in the recent ref.~\cite{Lv:2019cke} and in the references therein). Moreover, it should be emphasized that a fully systematic study of the value of $ma$ in the thermodynamic limit and very close to criticality would also require taking into account subleading corrections, which are neglected in eq.~(\ref{expected_correlator_form}): these include, for instance, short-distance corrections to $r$ that depend on the lattice geometry~\cite{Luscher:1995zz, Necco:2001xg}, contributions to $G(\tau,R)$ from additional periodic images of the zero-momentum operator, terms related to heavier excitations (which are exponentially suppressed with respect to the lightest mode), etc. While their discussion lies beyond the scope of this work, we remark that our present analysis of the $G(\tau,R)$ results provides a reliable and robust way to extract the mass of the lightest physical excitation of the theory for small but non-vanishing values of $t$.
\begin{table}[h]
\centering
\begin{tabular}{|r|l||r|l|}
\hline
\multicolumn{1}{|c|}{$L/a$} & \multicolumn{1}{|c||}{$ma$} & \multicolumn{1}{|c|}{$L/a$} & \multicolumn{1}{|c|}{$ma$} \\
 \hline \hline
$40$ & $0.1043(11) $ &  $80$ & $0.05105(35)$ \\
$48$ & $0.08373(60)$ &  $88$ & $0.04694(34)$ \\
$56$ & $0.06749(41)$ &  $96$ & $0.04005(12)$ \\
$64$ & $0.06099(45)$ & $104$ & $0.03302(19)$ \\
$72$ & $0.05829(32)$ &       &               \\
\hline
\end{tabular}
\caption{Results for the lightest mass contributing to the $G(\tau,R)$ correlator at $t \approx 0$, for different values of $L=R$.}\label{tab:m_results}
\end{table}
Following this strategy, we then evaluate the mass in units of the inverse lattice spacing $ma$ at fixed $t$ and for different lattice sizes, obtaining the results extrapolated to the thermodynamic limit through a fit to
\eq{large_volume_extrapolation_fit}
am(L)= am + \frac{ak_1}{L} .
\en
We note, in particular, that for $t \approx 0$ the thermodynamic-limit value of $am=-0.0058(37)$ is compatible with $0$ within less that two standard deviations. In addition, we also note that the systematic uncertainty associated with the choice of the functional form in eq.~(\ref{large_volume_extrapolation_fit}) can be (roughly) estimated by studying how the thermodynamic-limit value of $am$ varies, when a different functional form is chosen; if one includes an additional term $a^2 k_2/L^2$ on the right-hand side of eq.~(\ref{large_volume_extrapolation_fit}), the extrapolated value of $am$ in the large-volume limit changes to $am=-0.023(16)$, indicating that the compatibility of $am$ with zero is robust, and that the systematic uncertainty may be of a size similar to the statistical one.

\begin{figure*}[]
\centerline{\includegraphics[height=0.28\textheight]{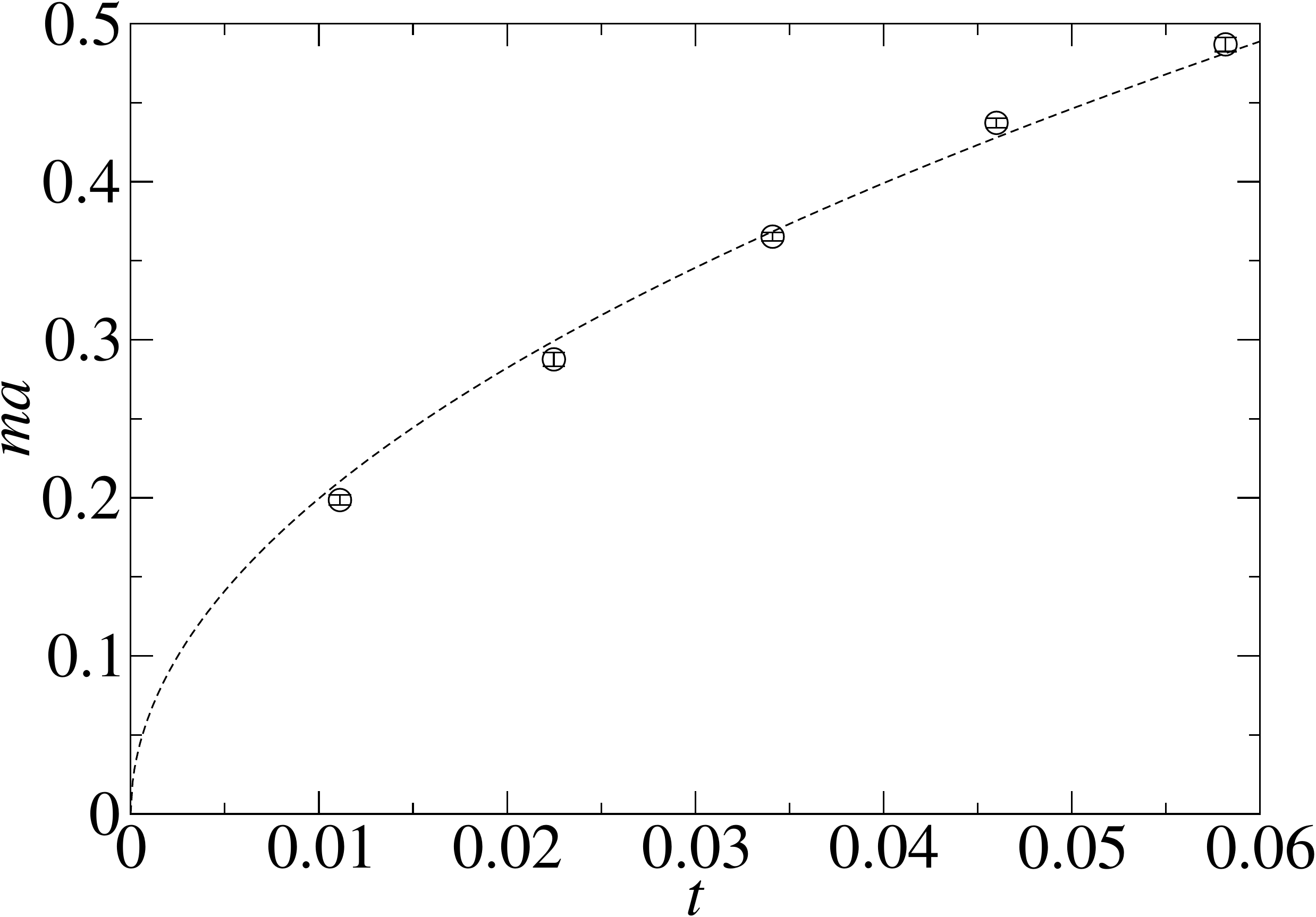}}
\caption{Mass values in units of the inverse lattice spacing at different reduced temperatures, extrapolated to the thermodynamic limit (circles), and their fit according to eq.~(\ref{m_versus_t}) (dashed line).} \label{fig:mplus_fit}
\end{figure*}

Neglecting logarithmic corrections, the masses extrapolated to the thermodynamic limit are expected to depend on $t$ as
\eq{m_versus_t}
ma = \frac{mt^\nu}{\Lambda_+},
\en
where $\Lambda_+$ is a constant with the dimensions of an energy, and a renormalization-group analysis predicts the critical exponent near the Gau{\ss}ian fixed point to be $\nu=1/2$. Fitting our results to eq.~(\ref{m_versus_t}), we obtain the amplitude value $m/\Lambda_+=1.995(24)$ and the results shown in figure~\ref{fig:mplus_fit}.

\begin{figure*}[]
\centerline{\includegraphics[height=0.275\textheight]{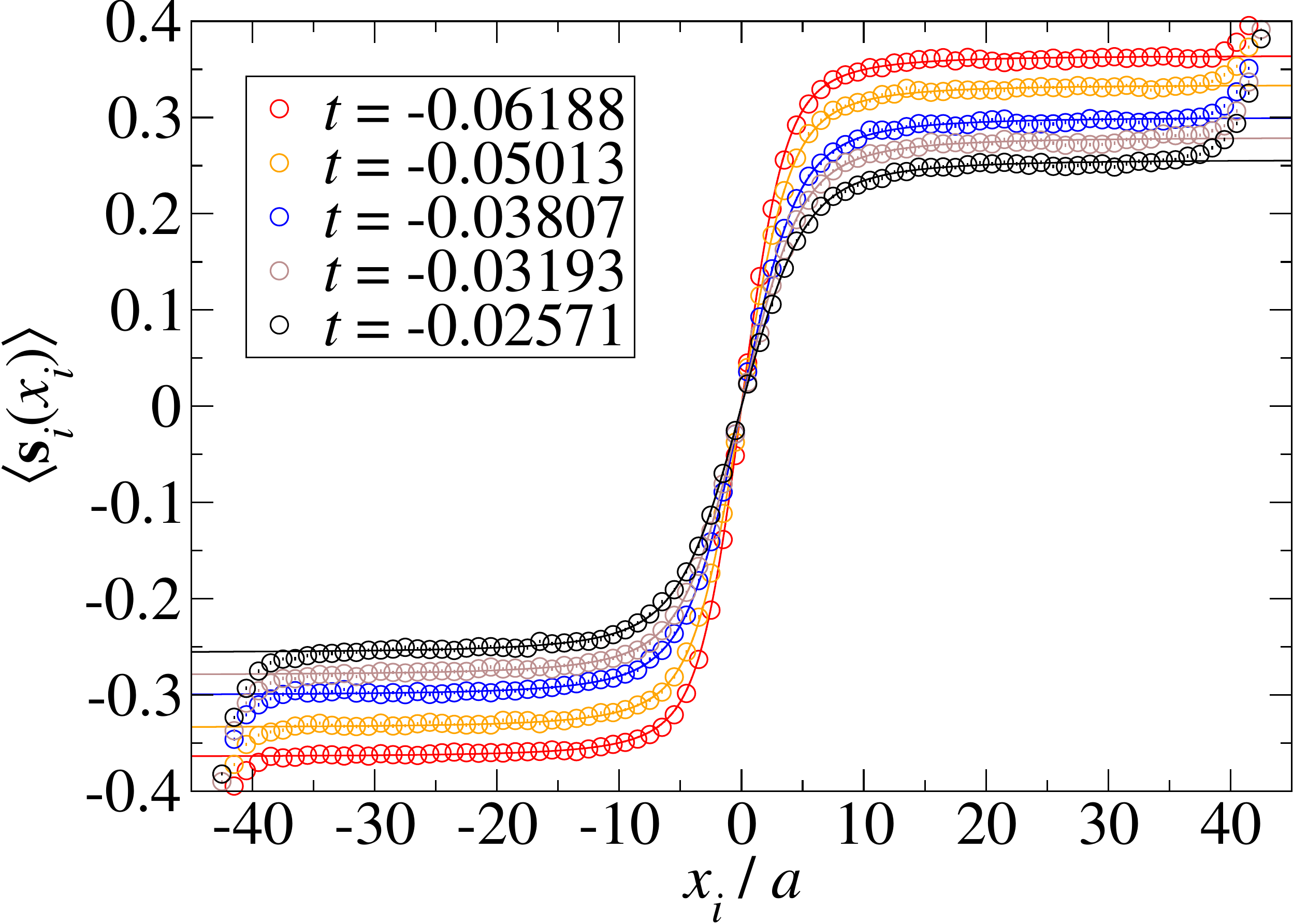} \hfill \includegraphics[height=0.275\textheight]{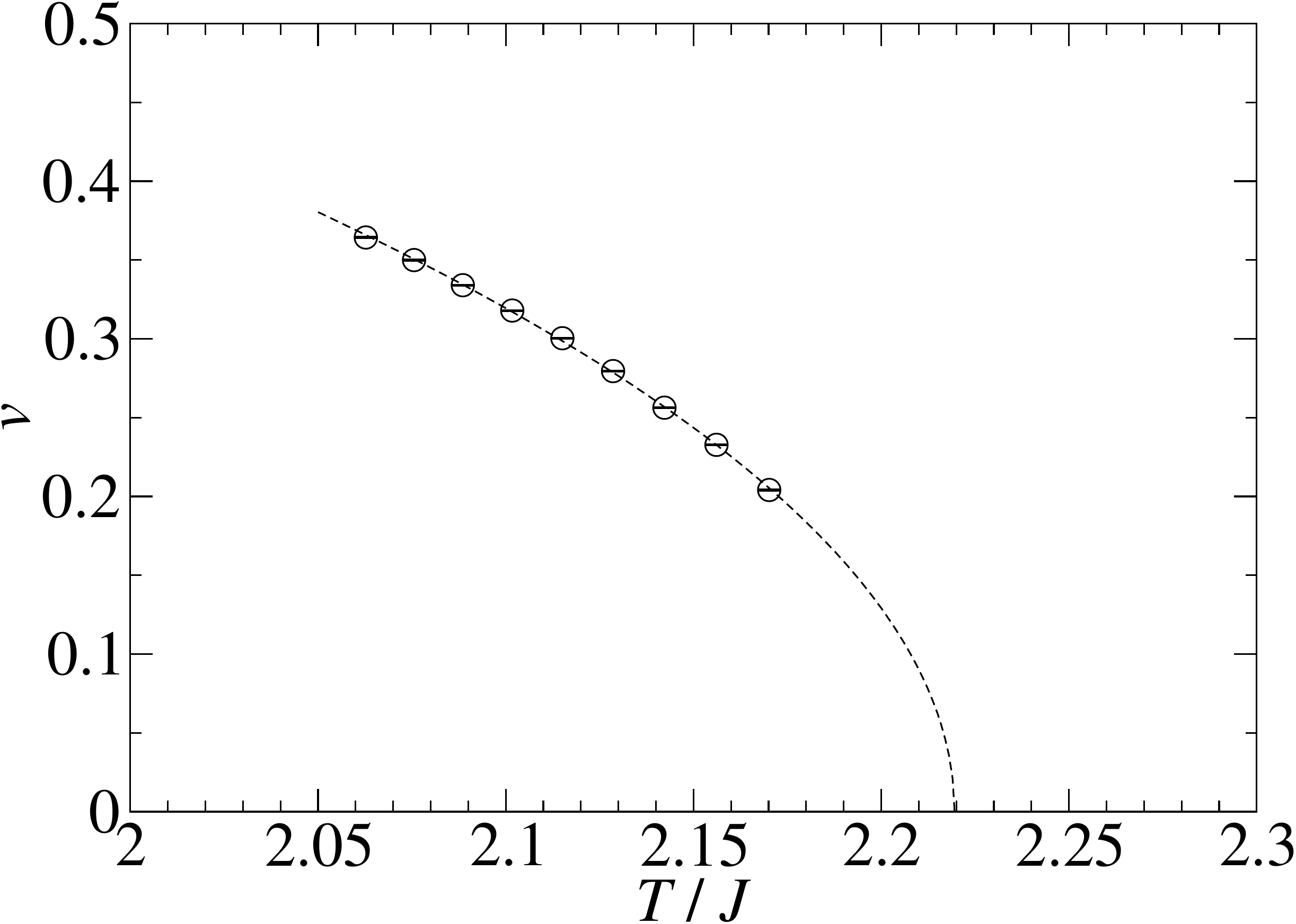}}
\caption{Left-hand-side panel: profile of the $i$-th component of the spin along the main axes through the center of the lattice, as a function of the $x_i$ coordinate (in units of the lattice spacing), in the presence of boundary conditions enforcing a topological excitation in the low-temperature phase. The plot shows the results obtained from our Monte~Carlo simulations on a lattice of spatial sizes $L=90a$ and Euclidean-time extent $R=20a$, denoted as circles of different colors, for different values of the reduced temperature $t$, and their comparison with the theoretical expectation according to eq.~(\ref{spin_profile_prediction}), which was derived in ref.~\cite{Delfino:2014rja} from quantum-field-theory arguments. Right-hand-side panel: the average spin value $v$ far from the defect core, obtained from fits to eq.~(\ref{spin_profile_prediction}), plotted against the temperature in units of the coupling, from different sets of simulations on lattices with $L=90a$ and $R=20a$ (including those displayed in the plot on the right-hand-side panel), and their fit to eq.~(\ref{v_fit}).}
\label{fig:spin_profile_vortex}
\end{figure*}

We now turn to the results of our Monte~Carlo simulations in the broken-symmetry phase at $T<\Tc$. In this case, we investigated the spin profile and energy density in the presence of a topological defect induced by the boundary conditions of the system, as described above. The left-hand-side panel of figure~\ref{fig:spin_profile_vortex} shows the spin profile along lines parallel to the three main spatial axes, and touching the edges of the cube at the center ¨of the lattice: more precisely, the plot displays the average value of the $i$-th component of the spin with the boundary conditions enforcing a topological defect, as a function of the coordinate of the $i$-th axis (in units of the lattice spacing). This quantity is averaged over the three directions, and is shown for a lattice of spatial sizes $L=90a$ and extent $R=20a$ in the Euclidean-time direction, and for different values of the reduced temperature $t$ (denoted by symbols of different colors). Our Monte~Carlo results are compared with the analytical predictions, denoted by solid curves, derived in ref.~\cite{Delfino:2014rja}:
\eq{spin_profile_prediction}
\langle \s_i (x_i) \rangle = v\left[ \left( 1-\frac{1}{2z^2} \right)\erf(z) + \frac{\exp\left(-z^2\right)}{\sqrt{\pi}z}\right],
\en
where $z=x_i\sqrt{2M/R}$, with $M$, which, according to ref.~\cite{Delfino:2014rja}, would represent the mass associated with the topological defect, and $v$, the asymptotic value of $\s_i$ at large distances from the defect core, as fit parameters. All fits are done in the range $-25 \le x_i/a \le 25$ to avoid systematic effects due to the boundaries of the system. We find excellent agreement between the theoretical curves and the numerical results, as shown in table~\ref{tab:spin_profile_fits}. 

\begin{table}[h]
\centering
\begin{tabular}{|l|l|l|l|}
\hline
\multicolumn{1}{|c|}{$t$} & \multicolumn{1}{|c|}{$v$} & \multicolumn{1}{|c|}{$Ma$} & \multicolumn{1}{|c|}{$\redchisq$} \\
 \hline \hline
$-0.01302$ & $0.2041(4)$   & $0.425(7)$  & $0.89$ \\              
$-0.01941$ & $0.2327(4)$   & $0.488(7)$  & $0.83$ \\             
$-0.02571$ & $0.2563(4)$   & $0.582(8)$  & $1.01$ \\              
$-0.03193$ & $0.27948(35)$ & $0.682(8)$  & $0.88$ \\             
$-0.03807$ & $0.30036(38)$ & $0.752(9)$  & $1.10$ \\              
$-0.04414$ & $0.31790(33)$ & $0.888(10)$ & $0.92$ \\              
$-0.05013$ & $0.33400(37)$ & $0.996(12)$ & $1.24$ \\              
$-0.05604$ & $0.34989(35)$ & $1.117(13)$ & $1.21$ \\              
$-0.06188$ & $0.3643(4)$   & $1.246(17)$ & $1.78$ \\  
\hline
\end{tabular}
\caption{Results for the fits of our numerical results for the spin profile, from simulations with $L/a=90$ and $R/a=20$, to eq.~(\ref{spin_profile_prediction}).}
\label{tab:spin_profile_fits}
\end{table}             

In addition, we also observe that the values for $v$ extracted from these fits exhibit the expected scaling when the temperature approaches $\Tc$: in the infinite-volume limit, the modulus of the magnetization is predicted to scale as $v \propto (-t)^\beta $, where (neglecting logarithmic corrections) the critical exponent is expected to take its Gau{\ss}ian value $\beta=1/2$. This is indeed confirmed by our results for $v$, which can be successfully fitted by
\eq{v_fit}
v = A_v \sqrt{1-\frac{T}{\Tc}},
\en
with $A_v=1.376(5)$ and $\Tc/J=2.2195(7)$, as shown on the right-hand side of figure~\ref{fig:spin_profile_vortex}. The slight mismatch between the fitted and the actual value of critical temperature provides an indication of the impact of finite-volume effects and logarithmic corrections on the critical exponent (that were neglected in this fit): their combined effect is below $1\%$. We also note that the precision of our results for $v$ is sufficient to rule out different values of the critical exponent. For example, fitting our data to a linear (in $T-\Tc$) form instead of eq.~(\ref{v_fit}) yields a reduced $\chi^2$ larger than $200$.

\begin{figure*}[]
\centerline{\includegraphics[height=0.28\textheight]{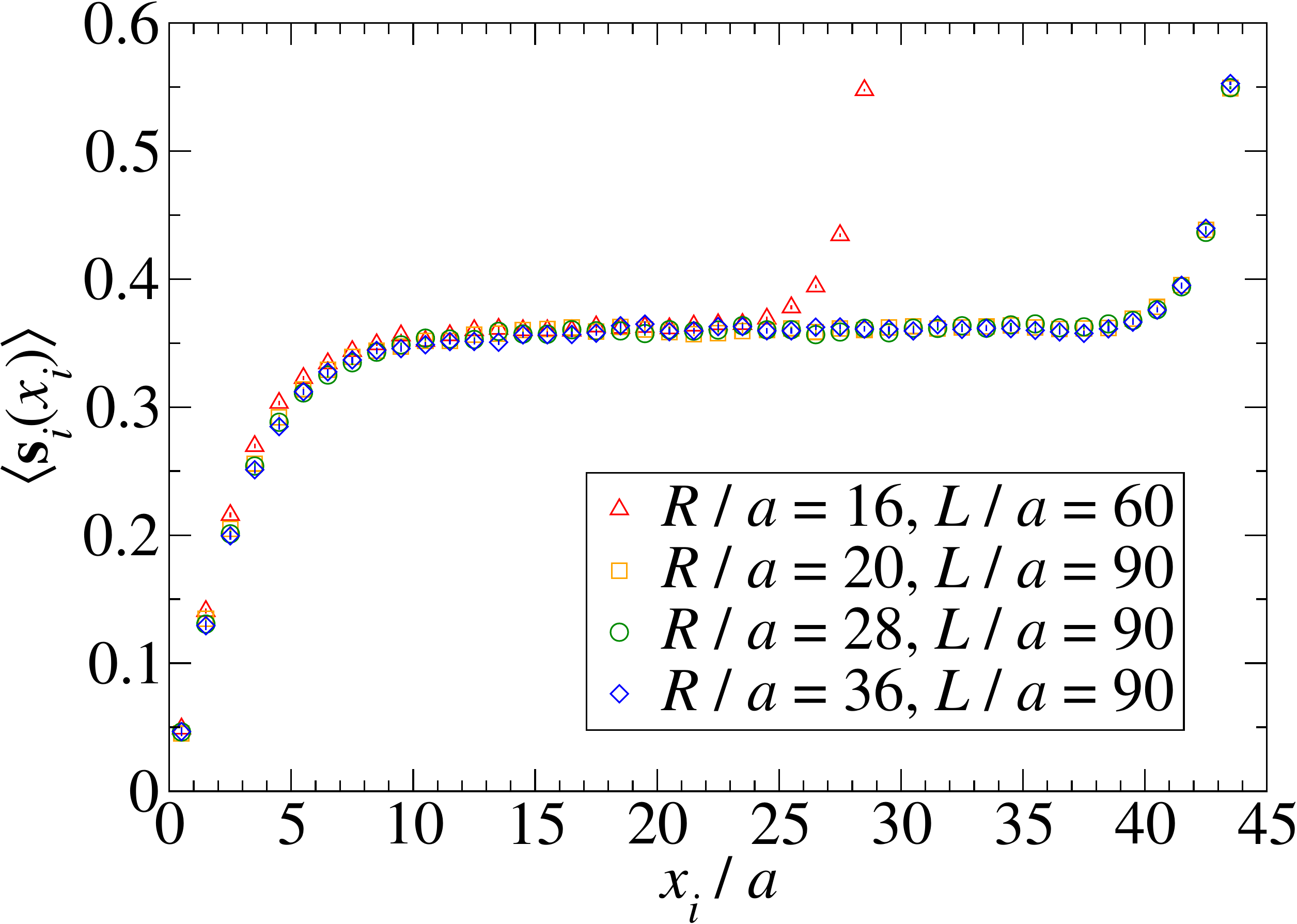}\hfill \includegraphics[height=0.28\textheight]{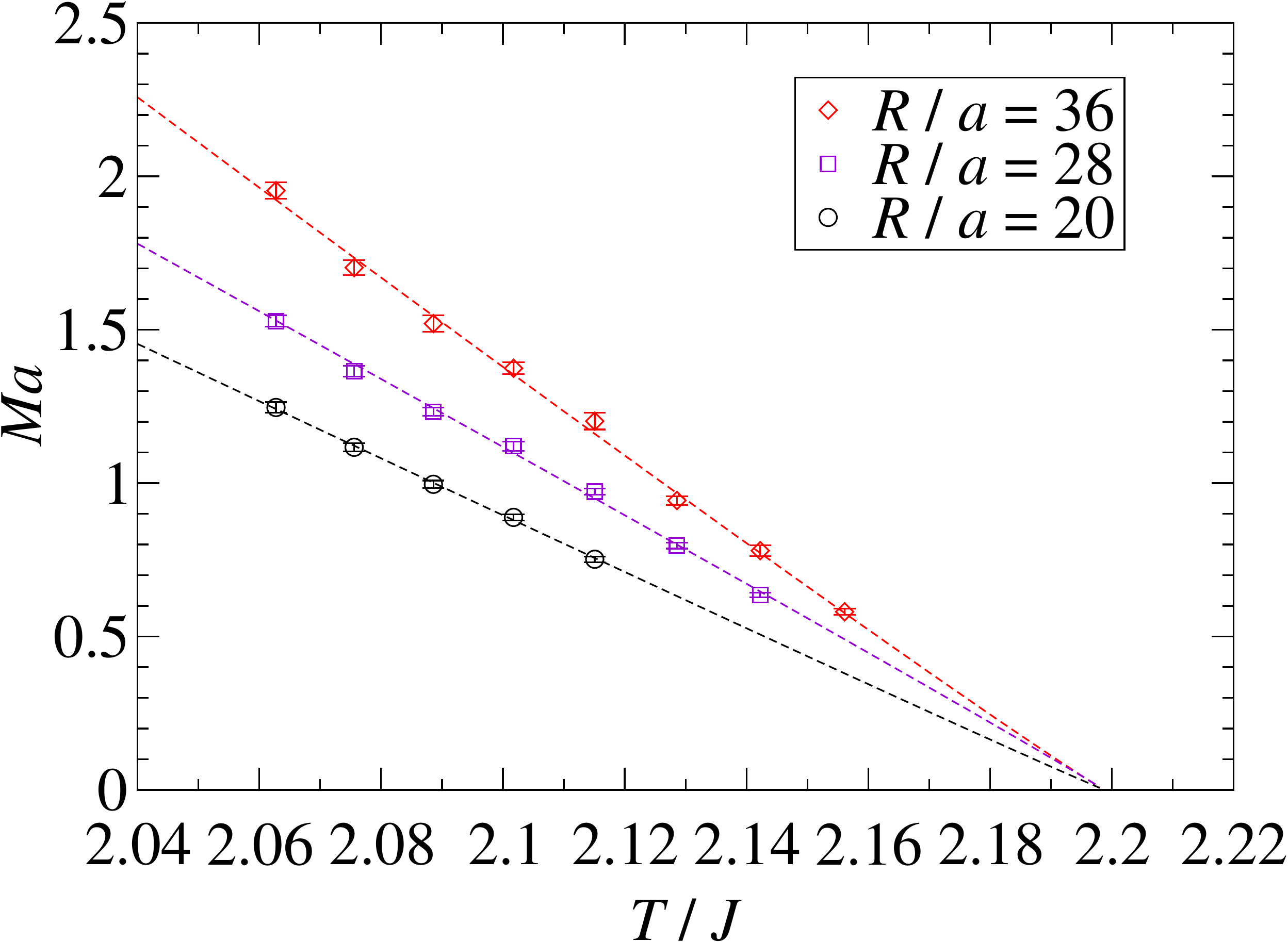}}
\caption{Left-hand-side panel: dependence of the spin profile $\langle \s_i (x_i) \rangle$ on the spatial coordinate $x_i/a$, as measured with respect to the center of the system, at a fixed temperature and for different values of the Euclidean-time extent of the lattice. All results shown in this plot were obtained from simulations with $T/J=2.062725$. Right-hand-side panel: temperature dependence of the $Ma$ parameter, extracted from the fits of our numerical data to eq.~(\ref{spin_profile_prediction}), for lattices with different values of $R/a$, and their fits to eq.~(\ref{twonu_fit}).} \label{fig:saturation}
\end{figure*}

It is interesting to study the scaling of the fitted value for $Ma$, as a function of the parameters of the theory, $L$, $R$ and $t$. The analytical calculations presented in ref.~\cite{Delfino:2014rja} are done in the thermodynamic limit, hence we restrict our analysis to lattices whose spatial volume $L^3$ is sufficiently large, in order to suppress finite-volume effects. More quantitatively, $L$ has to be much larger than the other length scales of the theory, i.e. the inverse of $M$ (implying $1 \ll ML$) and the lattice spacing ($1 \ll L/a$). Both inequalities are satisfied in our data samples. The dependence of $Ma$ on $R$, the lattice extent in the Euclidean-time direction, is more subtle. According to the interpretation of the topological field configuration as a particle excitation~\cite{Delfino:2014rja}, $M$ should be interpreted as the mass of the particle, and, as such, should be independent from $R$ (possibly up to discretization effects, for values of $R$ comparable with $a$). Our fit results, however, indicate that the dimensionless parameter $Ma$ extracted from the fits scales approximately proportionally to $R$, which questions the interpretation of the topological excitation as a physical particle. This is clearly revealed by the values of the spin profile $\langle \s_i (x_i) \rangle$ computed numerically at fixed temperature and for different values of $R$, an example of which (for $T/J=2.062725$) is shown in the plot on the left-hand-side panel of figure~\ref{fig:saturation}, which only shows results at $x_i/a \ge 0$: the data obtained on lattices with Euclidean-time extent $R/a=16$, $20$, $28$, and $36$, and spatial sizes $L\gtrsim 3 R$ collapse on the same curve (except for the points close to the boundaries of the lattice with $L=60a$), which can be described well by eq.~(\ref{spin_profile_prediction}). In turn, since the latter depends on $M$ only through the $M/R$ ratio, it follows that $M$ is proportional to $R$, or, equivalently, that the topological excitation is characterized by an approximately constant $\mu=M/R$. Note that, if this is the case, i.e. if $\mu$ should be thought of as a physical quantity, then in a set of simulations at fixed $R/a$ and at different temperatures, the $aM$ parameter extracted from the fits should scale as $a \mu R = \mu (R/a) a^2 \propto |t|^{2\nu}$. Remarkably, this behavior is indeed seen in our numerical data: as an example, the plot on the right-hand side of figure~\ref{fig:saturation} shows the results for $aM$ obtained from lattices with $R/a=36$ (red diamonds), $R/a=28$ (violet squares), and $R/a=20$ (black circles), at various temperatures (and for $L/a=90$). The data corresponding to each value of $R/a$ are fitted to the form
\eq{twonu_fit}
aM = A_M \left(1 - \frac{T}{\Tc}\right)^E
\en
(with the critical temperature fixed to its value computed in ref.~\cite{Lv:2019cke}), and, for all three cases, we always find results for the $E$ exponent very close to $1$, i.e. twice the value of $\nu$ predicted in the Gau{\ss}ian approximation. The fitted curves are shown as dashed lines in the right-hand-side plot of figure~\ref{fig:saturation}. Note that the dashed curves are not (or, more appropriately: are not constrained to be) straight lines. The fit results are reported in table~\ref{tab:twonu_fit_results}.

\begin{table}[h]
\centering
\begin{tabular}{|c|l|l|l|}
\hline
$R/a$ & \multicolumn{1}{|c|}{$A_M$} & \multicolumn{1}{|c|}{$E$} & \multicolumn{1}{|c|}{$\redchisq$} \\
 \hline \hline
$20$ & $21.3(1.5)$ & $1.021(23)$ & $0.50$ \\ 
$28$ & $23.4(1.8)$ & $0.981(24)$ & $2.34$ \\
$36$ & $34.6(2.3)$ & $1.039(21)$ & $1.76$ \\        
\hline
\end{tabular}
\caption{Results of the two-parameter fits of the data shown in the right-hand-side panel of figure~\ref{fig:saturation} to eq.~(\ref{twonu_fit}).}
\label{tab:twonu_fit_results}
\end{table}

\begin{figure*}[]
\centerline{\includegraphics[height=0.28\textheight]{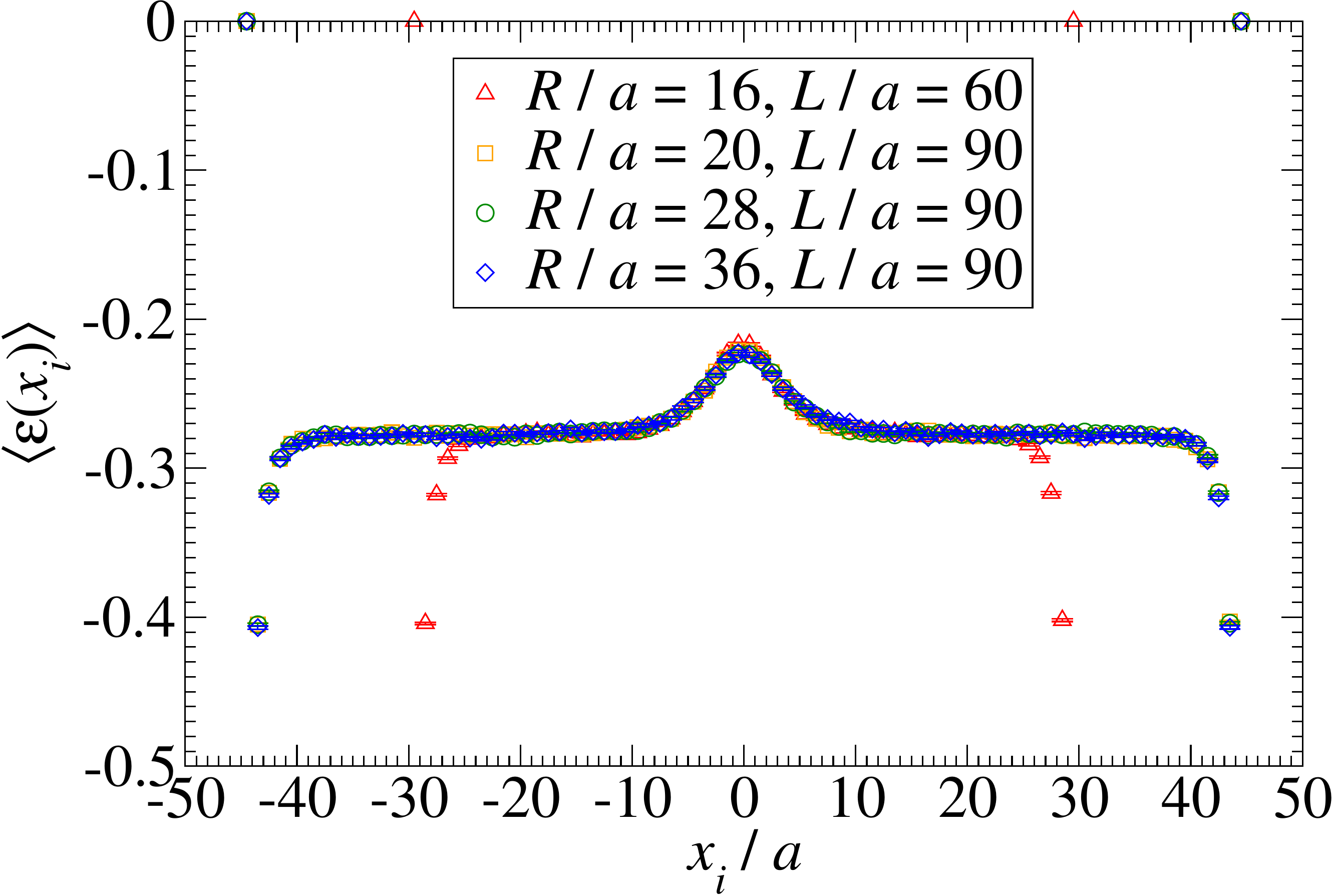} \hfill \includegraphics[height=0.28\textheight]{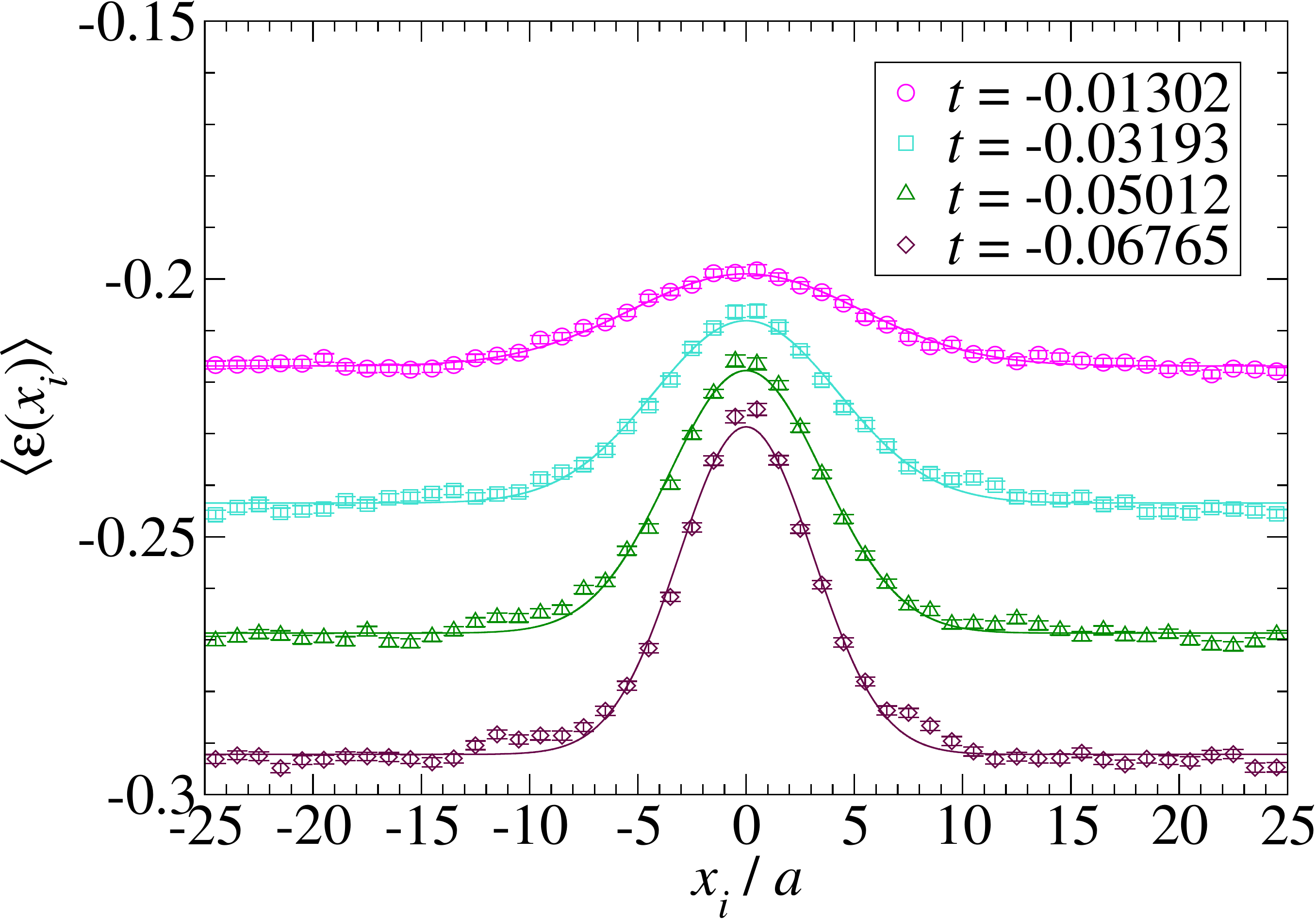}}
\caption{Left-hand-side panel: average energy density $\langle \varepsilon \rangle$ along the main spatial axes through the edges of a cube at the center of the lattice, as a function of the $x_i$ coordinate, in units of the lattice spacing, in the presence of a topological excitation in the broken-symmetry phase, as obtained from simulations at $t=-0.05604$. Right-hand-side panel: dependence of $\langle \varepsilon(x_i) \rangle$ on the reduced temperature, as evaluated from numerical simulations on lattices with Euclidean-time extent $R=20a$ and spatial volume $L^3=(90a)^3$. The solid curves are obtained from fits to eq.~(\ref{action_density_profile_prediction}), which was derived in ref.~\cite{Delfino:2014rja}.}
\label{fig:action_density_profile_vortex}
\end{figure*}

Finally, in figure~\ref{fig:action_density_profile_vortex} we present our results for $\varepsilon$, defined as the average (dimensionless) energy density along the main spatial axes through the edges of the cube at the center of the lattice. This quantity is plotted as a function of the coordinate $x_i$ along that axis (in units of $a$), averaging over the three spatial axes to enhance statistical precision. At any fixed temperature $T < \Tc$, in the infinite-volume limit $\varepsilon(x_i)$ is expected to tend to a spatially uniform value $C$ at sufficiently large distances. Like for the average spin profile, we observe that, for sufficiently large values of $L$ and $R$, also the average energy-density profile depends only on the temperature (for all points, except those close to the boundaries of the lattice), and, in particular, the numerical results obtained from Monte~Carlo simulations at the same $T$, but for different values of $R$, collapse on the same curve. An example of this behavior is shown in the plot on the left-hand side of figure~\ref{fig:action_density_profile_vortex}, displaying the results for $\langle \varepsilon (x_i) \rangle$ from simulations at $T/J=2.07557$, for $R/a=16$ (red triangles), $R/a=20$ (yellow squares), $R/a=28$ (green circles), and $R/a=36$ (blue diamonds); the spatial volume is $L^3=(90a)^3$, except for $R/a=16$, for which $L^3=(60a)^3$.

The plot on the right-hand side of figure~\ref{fig:action_density_profile_vortex} shows a comparison of our numerical results for the average energy-density profile to the theoretical prediction derived in ref.~\cite{Delfino:2014rja}, which is
\eq{action_density_profile_prediction}
\langle \varepsilon(x_i) \rangle = C+B M \sqrt{ \left( \frac{2Ma^2}{R}\right)^{3}} \exp\left( -\frac{2M}{R} x_i^2 \right).
\en
The symbols in the figure display data obtained from simulations on lattices with $R/a=20$ and $L/a=90$ at four different temperatures, focusing on the region around the center of the lattice, where the numerical results are fitted to eq.~(\ref{action_density_profile_prediction}), i.e. $-25 \le x_i/a \le 25$. As for the spin profile, the numerical results exhibit the expected qualitative features (for instance, the energy density has a peak at the center of the system, whose width becomes larger when $T \to \Tc^-$) and are in excellent quantitative agreement with the theoretical model.

\section{Discussion and conclusions}
\label{sec:discussion_and_conclusions}

To summarize, we studied the Heisenberg model on a four-dimensional Euclidean lattice, through massively parallel numerical simulations, based on state-of-the-art algorithms and run on high-performance-computing clusters. Even though $D=4$ is the upper critical dimension, the model reveals interesting dynamical features. As a benchmark, our study of correlators of zero-momentum spin operators in the disordered, high-temperature phase confirms the results recently reported in ref.~\cite{Lv:2019cke} (and based on the analysis of a different observable), yielding, in particular, full consistency with the critical temperature determined in that work. This may also be interpreted as indirect evidence for the non-trivial subleading finite-size correction terms discussed in that reference.

In the low-temperature phase, we implemented boundary conditions enforcing the existence of topologically non-trivial field configurations (which, in a continuum description, can be characterized by a non-zero ``winding number'' under the second homotopy group $\pi_2(S^2)$) and studied their propagation from an initial to a final Euclidean time. As discussed in ref.~\cite{Delfino:2014rja}, in the continuum limit the dynamical properties of this type of configurations can be studied with analytical tools from quantum field theory, using only general assumptions about the form factors relevant for different observables of the model as input. The predictions derived in this approach are supported by exact analytical solutions in two dimensions~\cite{Berg:1978sw, Delfino:2003yr} and by simulation results in three dimensions~\cite{Delfino:2018bff}. In the present work, for the first time, they have also been quantitatively confirmed in four Euclidean dimensions, to the high level of numerical precision that was possible to achieve through a large set of dedicated simulations, in a range of volumes spanning two orders of magnitude (the number of degrees of freedom on the largest lattices being approximately $2.34 \cdot 10^8$), and for a fine scan of temperatures close to $\Tc$. It is remarkable that the analytical approach advocated in ref.~\cite{Delfino:2014rja} has very strong predictive power, yielding quantitatively accurate expectations for various observables, from a very limited number of unknown parameters only, and for quite different models, based on continuous or discrete degrees of freedom, invariant under Abelian or non-Abelian symmetries, and defined in different spacetime dimensions.

Our simulation results also reveal that the scaling of the parameters (which could be interpreted as ``low-energy constants'') that appear in the quantum-field-theoretical description of the spin model for $T \to \Tc^-$ is non-trivial, and suggest that a straightforward interpretation of these topological configurations as particles is problematic. In particular, the data lead us to the conclusion that the parameter playing the r\^ole of the particle mass in ref.~\cite{Delfino:2014rja} is actually a quantity proportional to the ``duration'' (in Euclidean time, in this case) of its propagation. Our numerical evidence for this scaling is twofold: on the one hand, the results of simulations at fixed temperature and at different values of $R$ reveal that the characteristic widths of average spin and energy-density profiles do not grow with $\sqrt{R}$. Rather, their squared width appears to saturate at a given value of $R/M$, so that, even when the topological configuration is allowed to propagate for a longer Euclidean-time interval, it remains spatially localized. On the other hand, we showed that the $aM$ parameter, that we extracted from the fits, scales with the reduced temperature $t$ proportionally to $|t|^{2\nu}$, which indicates that the appropriate ``physical'' quantity that characterizes these objects is $\mu=M/R$, rather than $M$.

These results have interesting implications. In particular, the scaling properties observed in four dimensions can be compared and contrasted with those reported in an analogous numerical study in three dimensions~\cite{Delfino:2018bff}, which investigated the vortices in the XY model. In that work, it was found that the vortex has a well-defined mass in the continuum limit, which is approximately $2.1$~times larger than the mass of the lightest physical excitation in the disordered phase, and which, when expressed in units of the inverse lattice spacing, scales like $|t|^\nu$. Moreover, it was also pointed out that those findings provided indirect evidence that Derrick's argument~\cite{Derrick:1964ww} (a theorem in classical field theory implying that non-trivial, regular, static, soliton-like configurations in scalar field theory cannot exist in more than two spacetime dimensions, as their energy is unstable under scale transformations) may be violated at the quantum level. The possibility of an anomalous violation of Derrick's theorem is particularly intriguing, especially in the light of some arguments that have been recently put forward in axiomatic quantum field theory~\cite{Davies:2019wym}.

It should be noted that a violation of Derrick's theorem at the quantum level would entail far-reaching consequences, in many different areas of physics. As an example with particularly striking phenomenological implications, it is worth remarking that in the context of relativistic astrophysics, Derrick's theorem implies that bosonic stars consisting of particles that are the excitations of real scalar fields do not exist~\cite{Schunck:2003kk}. In fact, typical ways to evade Derrick's theorem consist in relaxing any of its assumptions, e.g. by introducing a local internal symmetry and gauge fields~\cite{Nielsen:1973cs, Polyakov:1974ek, tHooft:1974kcl} or replacing the real scalar field with a complex one, which may be charged under an unbroken, global, continuous internal symmetry and sustain \emph{stationary} (i.e. time-dependent and oscillating), rather than \emph{static}, configurations~\cite{Coleman:1985ki}. Finally, the possibility that Derrick's theorem may be violated in curved spacetimes has been recently studied in refs.~\cite{Alestas:2019wtw, Carloni:2019cyo}.

Alternatively, one can construct topologically stable, monopole-like field configurations of finite energy in physical systems characterized by an intrinsic cutoff scale: these include, in particular, those relevant for condensed-matter theory. It is, in fact, in this setting that the findings of our study may be particularly relevant: it is also worth pointing out that, while our calculations are carried out in a classical statistical mechanics setting, one could alternatively interpret this model as the lattice regularization of the corresponding quantum theory in three spatial dimensions~\cite{Montvay:1994cy}. The possibility that this theory supports artificial monopole-like objects~\cite{Volovik:1987lm, Castelnovo:2007qi, Milde:2013uo} may have important applications, given that synthetic magnetism is expected to provide a route towards quantum simulation~\cite{Lin:2009sm} and information communication~\cite{Hafezi:2011ro}. For a recent example of experimental work in this area, see, for instance, the study of manganese germanide reported in ref.~\cite{Kanazawa:2020doo}.

Finally, while the present numerical study has been carried out in a conventional, equilibrium Monte~Carlo setting, it would be interesting to generalize it to address aspects related to out-of-equilibrium dynamics. A possible way to do this, harnessing well-established mathematical theorems in non-equilibrium statistical mechanics~\cite{Jarzynski:1996oqb, Jarzynski:1997ef, Crooks:1997ne}, was recently discussed in ref.~\cite{Francesconi:2020fgi}, which presented a non-perturbative study of gauge theories subject to boundary conditions enforcing a non-zero minimum action field configuration, and determined the value of the physical running coupling, in a well-defined renormalization scheme~\cite{Symanzik:1981wd, Luscher:1992an}, by measuring the response of the system under a sequence of quantum quenches that ``deform'' the field configurations at the boundary. Repeating a similar type of computation for the setup considered in the present work could shed further light onto the properties of the topological configurations that we studied here, and onto their potential applications in condensed-matter systems.

\vskip1.0cm 
\noindent{\bf Acknowledgements}\\

The numerical simulations were run on machines of the Consorzio Interuniversitario per il Calcolo Automatico dell'Italia Nord Orientale (CINECA). We thank G.~Delfino for helpful discussions.

\bibliography{paper}

\providecommand{\href}[2]{#2}\begingroup\begin{thebibliography}{10}

\bibitem{Stanley:1968do}
H.~E. Stanley, \emph{{Dependence of Critical Properties on Dimensionality of
  Spins}}, \href{http://dx.doi.org/10.1103/PhysRevLett.20.589}{\emph{Phys. Rev.
  Lett.} {\bfseries 20} (Mar, 1968) 589--592}.

\bibitem{Froehlich:1981ib}
J.~Fr{\"o}hlich, B.~Simon and T.~Spencer, \emph{{Infrared bounds, phase
  transitions and continuous symmetry breaking}},
  \href{http://dx.doi.org/10.1007/BF01608557}{\emph{Commun. Math. Phys.}
  {\bfseries 50} (1976) 79--95}.

\bibitem{Kenna:2004cm}
R.~Kenna, \emph{{Finite size scaling for {$O(N)$} {$\phi^4$}-theory at the
  upper critical dimension}},
  \href{http://dx.doi.org/10.1016/j.nuclphysb.2004.05.012}{\emph{Nucl. Phys. B}
  {\bfseries 691} (2004) 292--304},
  [\href{https://arxiv.org/abs/hep-lat/0405023}{{\ttfamily hep-lat/0405023}}].

\bibitem{Bauerschmidt:2014sl}
R.~Bauerschmidt, D.~C. Brydges and G.~Slade, \emph{{Scaling limits and critical
  behaviour of the 4-dimensional n-component {$|\varphi|^4$} spin model}},
  \href{http://dx.doi.org/10.1007/s10955-014-1060-5}{\emph{Journal of
  Statistical Physics} {\bfseries 157} (Aug, 2014) 692--742}.

\bibitem{Delfino:2014rja}
G.~Delfino, \emph{{Order parameter profiles in presence of topological defect
  lines}}, \href{http://dx.doi.org/10.1088/1751-8113/47/13/132001}{\emph{J.
  Phys. A} {\bfseries 47} (2014) 132001},
  [\href{https://arxiv.org/abs/1401.2041}{{\ttfamily 1401.2041}}].

\bibitem{Berg:1978sw}
B.~Berg, M.~Karowski and P.~Weisz, \emph{{Construction of Green Functions from
  an Exact S Matrix}},
  \href{http://dx.doi.org/10.1103/PhysRevD.19.2477}{\emph{Phys. Rev. D}
  {\bfseries 19} (1979) 2477}.

\bibitem{Delfino:2003yr}
G.~Delfino, \emph{{Integrable field theory and critical phenomena: The Ising
  model in a magnetic field}},
  \href{http://dx.doi.org/10.1088/0305-4470/37/14/R01}{\emph{J. Phys. A}
  {\bfseries 37} (2004) R45},
  [\href{https://arxiv.org/abs/hep-th/0312119}{{\ttfamily hep-th/0312119}}].

\bibitem{Delfino:2018bff}
G.~Delfino, W.~Selke and A.~Squarcini, \emph{{Vortex mass in the
  three-dimensional {$O(2)$} scalar theory}},
  \href{http://dx.doi.org/10.1103/PhysRevLett.122.050602}{\emph{Phys. Rev.
  Lett.} {\bfseries 122} (2019) 050602},
  [\href{https://arxiv.org/abs/1808.09276}{{\ttfamily 1808.09276}}].

\bibitem{Lv:2019cke}
J.-P. Lv, W.~Xu, Y.~Sun, K.~Chen and Y.~Deng, \emph{{Finite-size Scaling of
  O({$n$}) Systems at the Upper Critical Dimensionality}},
  \href{http://dx.doi.org/10.1093/nsr/nwaa212}{\emph{National Science Review}
  (08, 2020) }, [\href{https://arxiv.org/abs/1909.10347}{{\ttfamily
  1909.10347}}].

\bibitem{Creutz:1980zw}
M.~Creutz, \emph{{Monte Carlo Study of Quantized SU(2) Gauge Theory}},
  \href{http://dx.doi.org/10.1103/PhysRevD.21.2308}{\emph{Phys. Rev.}
  {\bfseries D21} (1980) 2308--2315}.

\bibitem{Miyatake:1986ot}
Y.~Miyatake, M.~Yamamoto, J.~J. Kim, M.~Toyonaga and O.~Nagai, \emph{{On the
  implementation of the 'heat bath' algorithms for Monte Carlo simulations of
  classical Heisenberg spin systems}},
  \href{http://dx.doi.org/10.1088/0022-3719/19/14/020}{\emph{Journal of Physics
  C: Solid State Physics} {\bfseries 19} (1986) 2539--2546}.

\bibitem{Berg:2004fd}
B.~A. Berg, \emph{{Introduction to Markov chain Monte Carlo simulations and
  their statistical analysis}},
  \href{https://arxiv.org/abs/cond-mat/0410490}{{\ttfamily cond-mat/0410490}}.

\bibitem{Adler:1981sn}
S.~L. Adler, \emph{{An Overrelaxation Method for the Monte Carlo Evaluationvof
  the Partition Function for Multiquadratic Actions}},
  \href{http://dx.doi.org/10.1103/PhysRevD.23.2901}{\emph{Phys. Rev.}
  {\bfseries D23} (1981) 2901}.

\bibitem{Wolff:1988uh}
U.~Wolff, \emph{{Collective Monte Carlo Updating for Spin Systems}},
  \href{http://dx.doi.org/10.1103/PhysRevLett.62.361}{\emph{Phys. Rev. Lett.}
  {\bfseries 62} (1989) 361}.

\bibitem{Luscher:1995zz}
M.~L{\"u}scher and P.~Weisz, \emph{{Coordinate space methods for the evaluation
  of Feynman diagrams in lattice field theories}},
  \href{http://dx.doi.org/10.1016/0550-3213(95)00185-U}{\emph{Nucl. Phys.}
  {\bfseries B445} (1995) 429--450},
  [\href{https://arxiv.org/abs/hep-lat/9502017}{{\ttfamily hep-lat/9502017}}].

\bibitem{Necco:2001xg}
S.~Necco and R.~Sommer, \emph{{The N(f) = 0 heavy quark potential from short to
  intermediate distances}},
  \href{http://dx.doi.org/10.1016/S0550-3213(01)00582-X}{\emph{Nucl. Phys.}
  {\bfseries B622} (2002) 328--346},
  [\href{https://arxiv.org/abs/hep-lat/0108008}{{\ttfamily hep-lat/0108008}}].

\bibitem{Derrick:1964ww}
G.~Derrick, \emph{{Comments on nonlinear wave equations as models for
  elementary particles}}, \href{http://dx.doi.org/10.1063/1.1704233}{\emph{J.
  Math. Phys.} {\bfseries 5} (1964) 1252--1254}.

\bibitem{Davies:2019wym}
D.~Davies, \emph{{Quantum Solitons in any Dimension: Derrick's Theorem v.
  AQFT}},  \href{https://arxiv.org/abs/1907.10616}{{\ttfamily 1907.10616}}.

\bibitem{Schunck:2003kk}
F.~E. Schunck and E.~W. Mielke, \emph{{General relativistic boson stars}},
  \href{http://dx.doi.org/10.1088/0264-9381/20/20/201}{\emph{Class. Quant.
  Grav.} {\bfseries 20} (2003) R301--R356},
  [\href{https://arxiv.org/abs/0801.0307}{{\ttfamily 0801.0307}}].

\bibitem{Nielsen:1973cs}
H.~B. Nielsen and P.~Olesen, \emph{{Vortex Line Models for Dual Strings}},
  \href{http://dx.doi.org/10.1016/0550-3213(73)90350-7}{\emph{Nucl. Phys.}
  {\bfseries B61} (1973) 45--61}.

\bibitem{Polyakov:1974ek}
A.~M. Polyakov, \emph{{Particle Spectrum in the Quantum Field Theory}},
  {\emph{JETP Lett.} {\bfseries 20} (1974) 194--195}.

\bibitem{tHooft:1974kcl}
G.~'t~Hooft, \emph{{Magnetic Monopoles in Unified Gauge Theories}},
  \href{http://dx.doi.org/10.1016/0550-3213(74)90486-6}{\emph{Nucl. Phys. B}
  {\bfseries 79} (1974) 276--284}.

\bibitem{Coleman:1985ki}
S.~R. Coleman, \emph{{Q Balls}},
  \href{http://dx.doi.org/10.1016/0550-3213(86)90520-1}{\emph{Nucl. Phys. B}
  {\bfseries 262} (1985) 263}. [Erratum: Nucl.Phys.B 269, 744 (1986)].

\bibitem{Alestas:2019wtw}
G.~Alestas and L.~Perivolaropoulos, \emph{{Evading Derrick's theorem in curved
  space: Static metastable spherical domain wall}},
  \href{http://dx.doi.org/10.1103/PhysRevD.99.064026}{\emph{Phys. Rev. D}
  {\bfseries 99} (2019) 064026},
  [\href{https://arxiv.org/abs/1901.06659}{{\ttfamily 1901.06659}}].

\bibitem{Carloni:2019cyo}
S.~Carloni and J.~L. Rosa, \emph{{Derrick's theorem in curved spacetime}},
  \href{http://dx.doi.org/10.1103/PhysRevD.100.025014}{\emph{Phys. Rev. D}
  {\bfseries 100} (2019) 025014},
  [\href{https://arxiv.org/abs/1906.00702}{{\ttfamily 1906.00702}}].

\bibitem{Montvay:1994cy}
I.~Montvay and G.~M{\"u}nster, \emph{{Quantum fields on a lattice}}.
\newblock Cambridge Monographs on Mathematical Physics. Cambridge University
  Press, 3, 1997,
  \href{http://dx.doi.org/10.1017/CBO9780511470783}{10.1017/CBO9780511470783}.

\bibitem{Volovik:1987lm}
G.~E. Volovik, \emph{{Linear momentum in ferromagnets}},
  \href{http://dx.doi.org/10.1088/0022-3719/20/7/003}{\emph{Journal of Physics
  C: Solid State Physics} {\bfseries 20} (1987) L83--L87}.

\bibitem{Castelnovo:2007qi}
C.~Castelnovo, R.~Moessner and S.~L. Sondhi, \emph{{Magnetic monopoles in spin
  ice}}, \href{http://dx.doi.org/10.1038/nature06433}{\emph{Nature} {\bfseries
  451N7174} (2008) 42--45}, [\href{https://arxiv.org/abs/0710.5515}{{\ttfamily
  0710.5515}}].

\bibitem{Milde:2013uo}
P.~Milde, D.~K{\"o}hler, J.~J.~Seidel, L.~M. Eng, A.~Bauer, A.~Chacon et~al.,
  \emph{{Unwinding of a Skyrmion Lattice by Magnetic Monopoles}},
  \href{http://dx.doi.org/10.1126/science.1234657}{\emph{Science} {\bfseries
  31} (2013) 1076--1080}.

\bibitem{Lin:2009sm}
Y.-J. Lin, R.~L. Compton, K.~Jim{\'e}nez-Garc{\'{\i}}a, J.~V. Porto and I.~B.
  Spielman, \emph{{Synthetic magnetic fields for ultracold neutral atoms}},
  \href{http://dx.doi.org/10.1038/nature08609}{\emph{Nature} {\bfseries 462}
  (2009) 628--632}, [\href{https://arxiv.org/abs/1007.0294}{{\ttfamily
  1007.0294}}].

\bibitem{Hafezi:2011ro}
M.~Hafezi, E.~A. Demler, M.~D. Lukin and J.~M. Taylor, \emph{{Robust optical
  delay lines with topological protection}},
  \href{http://dx.doi.org/10.1038/nphys2063}{\emph{Nature Physics} {\bfseries
  7} (2011) 907--912}, [\href{https://arxiv.org/abs/1102.3256}{{\ttfamily
  1102.3256}}].

\bibitem{Kanazawa:2020doo}
N.~Kanazawa, A.~Kitaori, J.~S. White, V.~Ukleev, H.~M. R{\o}nnow, A.~Tsukazaki
  et~al., \emph{{Direct Observation of the Statics and Dynamics of Emergent
  Magnetic Monopoles in a Chiral Magnet}},
  \href{http://dx.doi.org/10.1103/PhysRevLett.125.137202}{\emph{Phys. Rev.
  Lett.} {\bfseries 125} (2020) 137202}.

\bibitem{Jarzynski:1996oqb}
C.~Jarzynski, \emph{{Nonequilibrium Equality for Free Energy Differences}},
  \href{http://dx.doi.org/10.1103/PhysRevLett.78.2690}{\emph{Phys. Rev. Lett.}
  {\bfseries 78} (1997) 2690--2693},
  [\href{https://arxiv.org/abs/cond-mat/9610209}{{\ttfamily
  cond-mat/9610209}}].

\bibitem{Jarzynski:1997ef}
C.~Jarzynski, \emph{{Equilibrium free-energy differences from nonequilibrium
  measurements: A master-equation approach}},
  \href{http://dx.doi.org/10.1103/PhysRevE.56.5018}{\emph{Phys. Rev.}
  {\bfseries E56} (1997) 5018--5035},
  [\href{https://arxiv.org/abs/cond-mat/9707325}{{\ttfamily
  cond-mat/9707325}}].

\bibitem{Crooks:1997ne}
G.~E. Crooks, \emph{{Nonequilibrium Measurements of Free Energy Differences for
  Microscopically Reversible Markovian Systems}},
  \href{http://dx.doi.org/10.1023/A:1023208217925}{\emph{J. Stat. Phys.}
  {\bfseries 90} (1998) 1481--1487}.

\bibitem{Francesconi:2020fgi}
O.~Francesconi, M.~Panero and D.~Preti, \emph{{Strong coupling from
  non-equilibrium Monte~Carlo simulations}},
  \href{http://dx.doi.org/10.1007/JHEP07(2020)233}{\emph{JHEP} {\bfseries 07}
  (2020) 233}, [\href{https://arxiv.org/abs/2003.13734}{{\ttfamily
  2003.13734}}].

\bibitem{Symanzik:1981wd}
K.~Symanzik, \emph{{Schr{\"o}dinger Representation and Casimir Effect in
  Renormalizable Quantum Field Theory}},
  \href{http://dx.doi.org/10.1016/0550-3213(81)90482-X}{\emph{Nucl. Phys.}
  {\bfseries B190} (1981) 1}.

\bibitem{Luscher:1992an}
M.~L{\"u}scher, R.~Narayanan, P.~Weisz and U.~Wolff, \emph{{The Schr{\"o}dinger
  functional: A Renormalizable probe for non-Abelian gauge theories}},
  \href{http://dx.doi.org/10.1016/0550-3213(92)90466-O}{\emph{Nucl. Phys.}
  {\bfseries B384} (1992) 168--228},
  [\href{https://arxiv.org/abs/hep-lat/9207009}{{\ttfamily hep-lat/9207009}}].

\end{thebibliography}\endgroup
\end{document}